\documentclass[final,5p,twocolumn]{elsarticle}

\usepackage{graphicx}
\usepackage{epsfig}

\usepackage{amssymb}
\def\beq{\begin{equation}}
\def\eeq{\end{equation}}
\def\Vec#1{\mbox{\boldmath $#1$}}
\def\beqy{\begin{eqnarray}}
\def\eeqy{\end{eqnarray}}
\journal{Physics Letters}

\begin{document}

  \begin{frontmatter}

%% Title, authors and addresses

%% use the tnoteref command within \title for footnotes;
%% use the tnotetext command for theassociated footnote;
%% use the fnref command within \author or \address for footnotes;
%% use the fntext command for theassociated footnote;
%% use the corref command within \author for corresponding author footnotes;
%% use the cortext command for theassociated footnote;
%% use the ead command for the email address,
%% and the form \ead[url] for the home page:
%% \title{Title\tnoteref{label1}}
%% \tnotetext[label1]{}
%% \author{Name\corref{cor1}\fnref{label2}}
%% \ead{email address}
%% \ead[url]{home page}
%% \fntext[label2]{}
%% \cortext[cor1]{}
%% \address{Address\fnref{label3}}
%% \fntext[label3]{}

    \title{A Monte Carlo generator of nucleon configurations in
      complex nuclei including Nucleon-Nucleon correlations}

    \author[label1]{M. Alvioli}
    \author[label2]{H. J. Drescher}
    \author[label1]{M. Strikman}

    \address[label1]{104 Davey Lab, The Pennsylvania State University,
      University Park, PA 16803, USA}
    \address[label2]{Frankfurt Institute for Advanced Studies, Johann Wolfgang
      Goethe-Universit\"at Routh-Moufang-Str. 1, 60438 Frankfurt am Main, Germany}

    \begin{abstract}
We developed a Monte Carlo event generator for production of nucleon configurations
in complex nuclei consistently including effects of Nucleon-Nucleon (NN) correlations.
Our approach is based on the Metropolis search for configurations satisfying essential
constraints imposed by short- and long-range NN correlations, guided by the findings
of realistic calculations of one- and two-body densities for medium-heavy nuclei.
The produced event generator can be used for Monte Carlo (MC) studies of pA and AA
collisions.
We perform several tests of consistency of the code and comparison with previous
models, in the case of high energy proton-nucleus scattering on an event-by-event
basis, using nucleus configurations produced by our code and Glauber multiple
scattering theory both for the uncorrelated and the correlated configurations;
fluctuations of the average number of collisions are shown to be affected considerably
by the introduction of NN correlations in the target nucleus.
We also use the generator to estimate maximal possible gluon nuclear shadowing
in a simple geometric model.
    \end{abstract}

%    \begin{keyword}
      %% keywords here, in the form: keyword \sep keyword

      %% PACS codes here, in the form: \PACS code \sep code

      %% MSC codes here, in the form: \MSC code \sep code
      %% or \MSC[2008] code \sep code (2000 is the default)

%    \end{keyword}

  \end{frontmatter}
%% \linenumbers
\section{Introduction}
The structure of nuclei has been described for a long time by independent
particle models in which the nucleus is treated as a collection of fermions
freely moving about, this picture being suggested by the fact that nuclear
matter is a dilute system. Nevertheless the short range structure of nuclei
cannot be accurately described by this simplified picture, and essential
aspects of nuclei such as high momentum component of the nuclear wave function
are completely missed by the non-interacting model; configurations of
nucleons at short separations are known as short range correlations
(SRC)\cite{Higinbotham:2009zz,Subedi:2008zz}.
NN correlations have been recently unambiguously observed\cite{Shneor:2007tu}
in a series of dedicated experiments, in which high-momentum nucleons were
knocked out by high energy probes and their correlated partner were observed
with opposite momentum; the probability for a nucleon to belong to a SRC
pair was measured to be about 20\% in $^{12}C$ and 25\% in in heavy nuclei;
see Ref.\cite{Frankfurt:2008zv} for a review.
A theoretical description of SRC can be made in a satisfactory way using
nonrelativistic Hamiltonians containing two- and three-body realistic potentials.
It is currently possible to solve exactly the Schr\"odinger equation for light
nuclei while, for $A>12$, approximate methods must be considered, which have been
shown to produce a reasonable approximation to the basic ground state quantities
such as one- and two-body densities and momentum distributions.
These studies demonstrated that NN correlations produce non-negligible effects
in a variety of processes including those which are not specially probing SRC.
For example, significant effects are found for high energy total nucleon-nucleus
cross sections\cite{Alvioli:2008rw}, and it has thus been shown that they must be
taken into account in the theoretical calculation of such processes.
In this Letter we are addressing the challenging problem of including NN correlations
in the description of high energy proton-nucleus and nucleus-nucleus collisions.
It was shown in Ref.\cite{Baym:1995cz} that using correlated two-body densities for
the analysis of fluctuations of the mean number of participating nucleons in
proton-nucleus collisions significantly modifies the results, especially the tails
of the distribution, and it was pointed out the importance of implementing NN correlations
in the event generator for nucleon configurations in a way consistent with the single
nucleon nuclear density and use it for description of the heavy ion collisions on an
event-by-event basis.
Simulations of such processes are commonly performed using configurations of nucleons
as an input, consisting of nucleons spatial locations and isospins, generated assuming
the independent particle model; however, inclusion of NN correlations can provide a
more realistic description of initial states of nuclei and appears to be feasible
with a good accuracy.

The Letter is organized as follows; in section II, details of NN correlations are
discussed and the method adopted for generation of configurations is outlined;
we describe the generation of nuclear configurations within a MC approach
including central two-body correlations which are consistent with realistic
calculations of one- and two-body density distributions. In section III  we introduce
the probability distribution functions $P_N$ for the $N$-th order, $N=1,...,A$, interaction
of a projectile nucleon with a nucleus $A$, and we present results for fluctuations.
In section IV we use our event generator to calculate maximal possible shadowing for gluons
based on simple geometric considerations.
\section{NN correlations}
The aim of the present work is to develop a MC procedure for generating spatial nucleon
configurations to be used as an input for simulations of collisions involving heavy nuclei which
treats NN correlations in a realistic way.
The commonly used approach to this problem is taking a given number of nucleons
out of a Woods-Saxon density distribution, which describes the probability density
function of a nucleon to be located at a distance $r$ from the center of the system.
As a result, the $A$ nucleons are positioned independently from each other,
such that all two-particle (and higher) correlations are completely ignored;
this is justified by the assumption that inclusive quantities are not very
much dependent on these correlations. Alternatively one puts nucleons in consequently
one after another imposing the condition that the distance between nucleons should be
larger than some minimal one. This procedure, however, results in a wrong single nucleon
distribution of the nucleons and must be improved.

We argue that when considering event-by-event observables, NN correlations in the nuclear
wave function are relevant. This extends observation of\cite{Baym:1995cz} that NN correlations
significantly modify the variance of the distribution over the number of collisions.
In the  present work we analyze several different approximations for correlations.
A full implementation of realistic short range
correlations\cite{Alvioli:2005cz,Pandharipande:1979bv} in
the wave function $\Psi$ of a complex nucleus can be made using an independent
particle model wave function $\Phi$ and a proper correlation operator
$\hat{F}=\prod^A_{i<j}\hat{f}(r_{ij})$ as follows:
\beq
\label{corrop}
\Psi(\Vec{r}_1,...,\Vec{r}_A)\,=\,\hat{F}\,\Phi(\Vec{r}_1,...,\Vec{r}_A)\,,
\eeq
$\hat{f}$ being a set of state-dependent correlation functions
$\hat{f}(r_{ij})=\sum_n f^{(n)}(r_{ij})\hat{O}^{(n)}_{ij}$ obtained variationally
from a nonrelativistic Hamiltonian with realistic two- and three-body interactions;
$\hat{O}^{(n)}_{ij}$ is the spin-isospin operator appearing in the realistic
two-body potential, usually written in the form
$\hat{V}\,=\,\sum_{i<j}\,\hat{v}(r_{ij})\,=\sum_{i<j}\,\sum_n\,v^{(n)}(r_{ij})
\,\hat{O}^{(n)}_{ij}\,.$
In the present implementation of correlations, we did not consider the full realistic
set of correlations shown in Eq.(\ref{corrop}), which would imply using the full $|\Psi|^2$
as a probability function for the Metropolis random search. This would be numerically very
demanding, especially for heavy nuclei, as it would involve generating a full configuration
of $A$ nucleons.
Hence as a first step we will implement only  effective central correlations; the inclusion
of spin- and isospin-dependent configurations, which allows to account for most of the properties
of one- and two-body densities described in\cite{Alvioli:2005cz,Alvioli:2007zz,Schiavilla:2006xx}
and references therein, will be discussed later on.

A major problem with using a system of uncorrelated nucleons for the event generators
is the occurrence of hot spots: regions in space where two or more nucleons are allowed
to overlap and produce unrealistically high local densities. A simple excluded volume
model, in which nucleons are prevented from being closer than a fixed distance
of about the nucleon size $d=1$ $fm$ does not account for the realistic picture of nuclei,
since this approach artificially moves density from the center of the nucleus towards the
periphery, distorting the nuclear profile and altering the surface properties. The first
constraint in our generator is thus to preserve the one body density of the nucleus,
defined as follows:
\beqy
\label{rho1}\rho^{(1)}(\Vec{r}_1)&=&A\,\int \prod^A_{i=2}\,d\Vec{r}_i
\,\left|\Psi(\Vec{r}_1,...,\Vec{r}_A)\right|^2\,.
\eeqy
We suggest a method for generating the spatial configurations of nuclei with the following
essential features:
\textit{i)} the nucleons are distributed according to a given single-particle distribution;
\textit{ii)} each configuration embodies the NN pair correlations.
The resulting two-body densities are by far more realistic than the commonly used
independent particle model ones, or even those of the excluded volume approximation.
The outlined results can be achieved by first generating a large number of random nucleon
positions in a cube with given density ($0.17$ $nucleons/fm^3$). Next, a
Metropolis random search is performed, with a probability function given by the square of
the wave function; at this stage, we impose the constraint of the Woods-Saxon probability
distribution (Ref.\cite{Gyulassy:1994ew}),
choosing those nucleons which are relevant to the desired
one-body density. The procedure comes to an end when exactly A nucleons satisfy the constraints
at the same time.

It was shown in Refs.\cite{Alvioli:2005cz,Alvioli:2007zz} that SRC have sizable effects
on the single particle density (Eq.(\ref{rho1})) as well as on those quantities which
depend depend on the two-body density of the system, defined as
\beq
\label{rho2}\rho^{(2)}(\Vec{r}_1,\Vec{r}_2)\,=\,A(A-1)
\,\int \prod^A_{i=3}\,d\Vec{r}_i\,\left|\Psi(\Vec{r}_1,...,\Vec{r}_A)\right|^2,\,
\eeq
such as the two-body momentum distributions. In many theoretical studies, Eq.(\ref{rho2}) is
approximated as follows
\beq
\label{rho12}
\hspace{-0.5cm}\rho^{(2)}(\Vec{r}_1,\Vec{r}_2)\,=\,\frac{A-1}{A}\,\rho^{(1)}(\Vec{r}_1)\rho^{(1)}(\Vec{r}_2)
\,\left[1\,-\,C(\Vec{r}_1,\Vec{r}_2)\right]
\eeq
where $C(\Vec{r}_1,\Vec{r}_2)$ vanishes if NN and statistical correlations are disregarded.
We can compare the realistic results\cite{Alvioli:2005cz,Pieper:1992gr} for the radial two-body
density, defined as the integral of Eq.(\ref{rho12}) over the pair center of mass, $\Vec{R}$:
\beq
\label{rho12r}
\hspace{-0.2cm}\rho^{(2)}(r)\,=\,\int d\Vec{R}\,\rho^{(2)}\left(\Vec{r}_1=\Vec{R}+\frac{1}{2}\Vec{r},
\Vec{r}_2=\Vec{R}-\frac{1}{2}\Vec{r}\right)\,
\eeq
and the results from our MC generator for the same quantity. If we assume that
$C(\Vec{r}_1,\Vec{r}_2)$ depends only on the relative coordinate
$r=\left|\Vec{r}_1-\Vec{r}_2\right|$ (which holds true exactly for infinite nuclear
%------------------------------------------------------------------------------%
%------------------------------------------------------------------------------%
\begin{figure}[!ht]
\vskip -0.3cm
\centerline{\includegraphics[width=10.4cm]{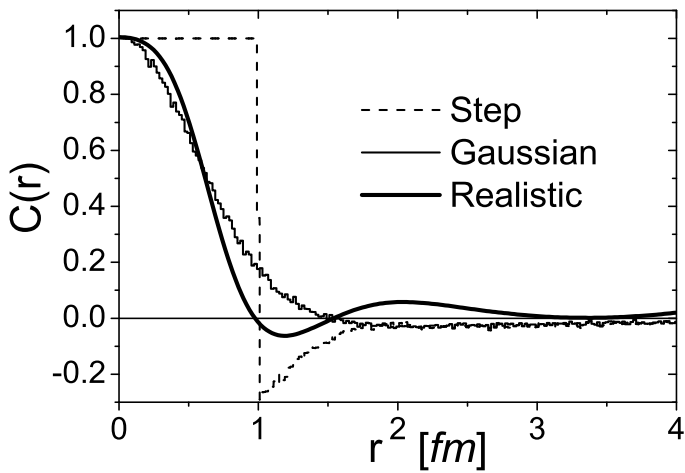}}
\vskip -0.3cm
\caption{
  The quantity of Eq.(\ref{c12r}), calculated for $^{16}O$.
  \textit{Dashed histogram}: step function (excluded volume) correlation;
  \textit{solid histogram}: Gaussian correlation;
  \textit{solid curve}: the realistic calculation of Ref.\cite{Alvioli:2005cz}.}
\label{Fig1}
\end{figure}
%------------------------------------------------------------------------------%
%------------------------------------------------------------------------------%
matter and which is a good approximation in finite nuclei, far enough from the edge
of the nucleus), the pair distribution function $C(r)$ can be obtained from the
correlated $\rho^{(2)}_C(r)$ and uncorrelated $\rho^{(2)}_U(r)$ two-body radial
densities:
\beq
\label{c12r}
C(r)\,=\,1\,-\,\rho^{(2)}_{C}(r)\,/\,\rho^{(2)}_U(r)\,.
\eeq
The quantities defined in Eq.($\ref{rho12r}$) and Eq.($\ref{c12r}$) were calculated
using the MC code using several approximations for the correlation function
$f(r)$ (see Eq.(\ref{corrop})). They are shown in Fig.\ref{Fig1}; the results of
Ref.\cite{Alvioli:2005cz} are also shown. Our procedure was to start with a central correlation
function $f(r)$ taken as a step function, i.e. vanishing wave function for $r<1$ $fm$; we then
used a Gaussian $f(r)=1-exp(-\beta\,r^2)$ in order to match better the realistic result. It is
known that the realistic central correlations has a small overshooting over unity, and it
is in general different from our Gaussian correlation; however, in realistic calculations of
finite nuclei\cite{Alvioli:2005cz,Pieper:1992gr} the peak in the radial two-body density is
mainly due to the existence of state-dependent, mainly tensor correlations, and three-body diagrams.
For this reason we prefer not to introduce an unrealistic central correlation to reproduce such
a peak and use a parameter $\beta=0.9$ $fm^{-2}$ in the Gaussian correlation which gives a healing
distance similar to the one exhibited by the realistic approach. This amounts to calculate the wave
function with a central correlation operator; the fully realistic implementation of the operator of
Eq.(\ref{corrop}), involving spin and isospin degrees of freedom, extremely computationally intensive,
is thus beyond the aim of the present contribution and will be performed elsewhere. In the present work
we randomly assigned isospin degrees of freedom for A nucleons, so that each generated configuration
%------------------------------------------------------------------------------%
%------------------------------------------------------------------------------%
\begin{figure}[!ht]
\centerline{\includegraphics[width=10.1cm]{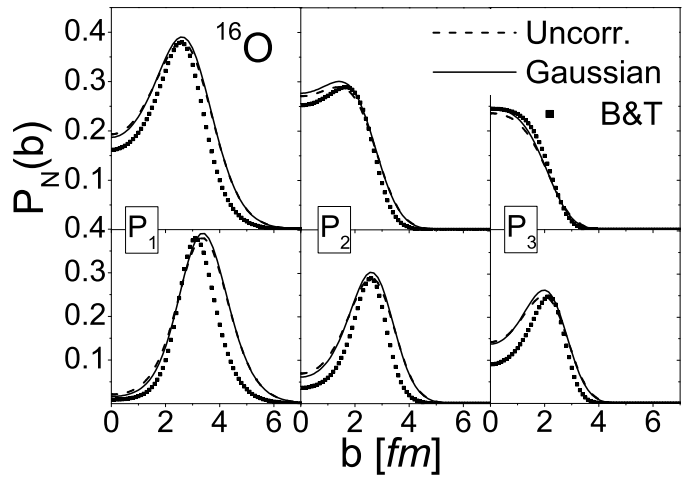}}
\vskip -0.0cm
\caption{The probabilities $P_N$, $N=1,2,3$, as a function of the impact
  parameter $b$ (see text), calculated for $\sqrt{s}=42$ $GeV$
  (\textit{top panels}) and for $\sqrt{s}=5500$ $GeV$ (LHC energies;
  \textit{lower panels}) on the $^{16}O$ nucleus.
  \textit{Squares}: analytic results obtained with the approximation of
  Eq.(\ref{thickness}), the Glauber approach with zero-range interaction of
  Bertocchi and Treleani, Ref.\cite{Bertocchi:1976bq};
  \textit{dashes}: uncorrelated result;
  \textit{solid lines}: Gaussian correlations.}
\label{Fig2}
\end{figure}
%------------------------------------------------------------------------------%
%------------------------------------------------------------------------------%
consists of A quads of numbers, containing three spatial coordinates and one, randomly assigned,
isospin variable; the correlated nuclear configurations are downloadable from the PSU physics
department web server at the address\cite{download}.
\section{Probability distribution functions for hadron-nucleus collisions}
In order to illustrate the present implementation of correlations, we have focused
on the $^{16}O$, $^{40}Ca$ and $^{208}Pb$ nuclei.

For a given configuration of nucleons, the probability of interaction with the $i$-th
nucleon for an incoming projectile with impact parameter $\Vec{b}$ is given by
\beq
\label{probint}
P(\Vec{b},\Vec{b}_i)\,=\,1\,-\,\left[1\,-\,\Gamma(\Vec{b}-\Vec{b}_i)\right]^2\,,
\eeq
the corresponding probability of no interaction being $1 - P(\Vec{b},\Vec{b}_i)$.
The $\Gamma$ function in Eq.(\ref{probint}) for high energy incident nucleons can
be parameterized as
\beq
\Gamma(\Vec{s})\,=\,\frac{\sigma_{NN}^{tot}}{4\pi B}\,e^{-s^2/2B}\,,
\eeq
where $\sigma_{NN}^{tot}$ is the total cross section of NN scattering and the
$t$ dependence of the cross section $d\sigma/dt \propto exp( B t)$, neglecting
small corrections due to the real part of the amplitude.
Let us define the probability of interaction with $N$ nucleons as a function
of the impact parameter as
\beqy
\label{probs}
&&\hspace{-1.3cm}P_N(\Vec{b})=\nonumber\\
&&\hspace{-1.3cm}\sum^N_{i_1,\dots,i_N}\,P(\Vec{b},\Vec{b}_{i_1})
\cdot\dots\cdot P(\Vec{b},\Vec{b}_{i_N})\hspace{-0.2cm}\prod^{A-N}_{j\neq i_1,\dots,i_N}
\hspace{-0.2cm}\left[1\,-\,P(\Vec{b},\Vec{b}_j)\right]\,.
\eeqy
%------------------------------------------------------------------------------%
%------------------------------------------------------------------------------%
\begin{figure}[!ht]
\vskip -0.6cm
\centerline{\includegraphics[width=10cm]{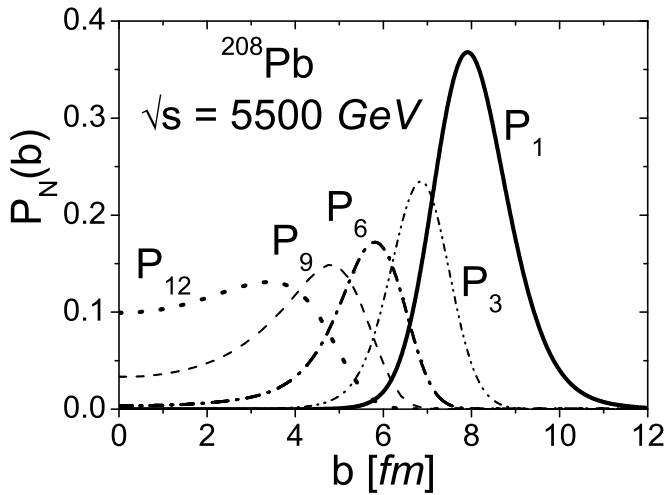}}
\vskip -0.3cm
\caption{The probabilities $P_N$, $N=1,3,6,9,12$, as a function of the
    impact parameter $b$, calculated for $\sqrt{s}=5500$ $GeV$,
    corresponding LHC energies, on the $^{208}Pb$ nucleus, using
    (gaussian) correlated configurations.}
\label{Fig3}
\end{figure}
%------------------------------------------------------------------------------%
%------------------------------------------------------------------------------%
The \textit{inelastic} cross section due to collisions with $N$ nucleons is then
given by
\beq
\label{siginN}
\sigma^{in}_N\,=\,\int d\Vec{b}\,P_N(\Vec{b})\,,
\eeq
and the \textit{total inelastic} cross section is
$\sigma^{in}_{NA}\,=\,\sum^A_i\,\sigma^{in}_i$,
and it can also be calculated as
\beq
\sigma^{in}_{NA}\,=\,\sigma^{tot}_{NA}\,-\,\sigma^{el}_{NA}\,-\,\sigma^{qe}_{NA}\,,
\eeq
$\sigma^{el}_{NA}$ and $\sigma^{qe}_{NA}$ being the elastic and quasi-elastic cross
section, respectively; this calculation is being performed with realistic wave functions
(Ref.\cite{alv04}) and a comparison with the present approach will be presented
elsewhere.

We can evaluate the probabilities given by Eq.(\ref{probs}) to any order in $N\le A$
as a function of $b = |\Vec{b}|$. The results for $A=16,208$ are shown in
Figs.\ref{Fig2}, \ref{Fig3}.
The number of configurations used in all the calculations is large enough to produce
negligible statistical errors in the results, typically over 100 thousands configurations.
All the results have been obtained using parameters of the NN amplitude corresponding to
$\sqrt{s}=42$ $GeV/c$ ($\sigma_{tot}=41.6$ $mb$, $B=12.6$ $GeV^{-2}$), except the lower
panel of Fig.\ref{Fig2} for $^{16}O$, where $\sqrt{s}=5500$ $GeV/c$ ($\sigma_{tot}=94.8$
$mb$, $B=17.3$ $GeV^{-2}$) is chosen, corresponding to LHC energies, to be compared with the
top panel of the same figure, and in Fig.\ref{Fig3} for $^{208}Pb$ where $\sqrt{s}=5500$ $GeV/c$
as well.

We have checked the independent particle model results with the analytic approximation
of Ref.\cite{Bertocchi:1976bq}, where the probability distributions are given by:
\beq
\label{thickness}
\hspace{-0.5cm}P_N(\Vec{b})=\frac{A!}{(A-N)! N!}\left(\sigma^{in}T(\Vec{b})\right)^N
\hspace{-0.1cm}\left[1-\sigma^{in}T(\Vec{b})\right]^{(A-N)}\hspace{-0.2cm},\,
\eeq
where $T(\Vec{b})\,=\,\int^\infty_{-\infty}dz\,\rho(\Vec{b},z)$, which neglects the
finite radius of NN interaction. The comparison is made for
%------------------------------------------------------------------------------%
%------------------------------------------------------------------------------%
\begin{figure}[!ht]
\vskip -0.0cm
\centerline{
    \includegraphics[width=10cm]{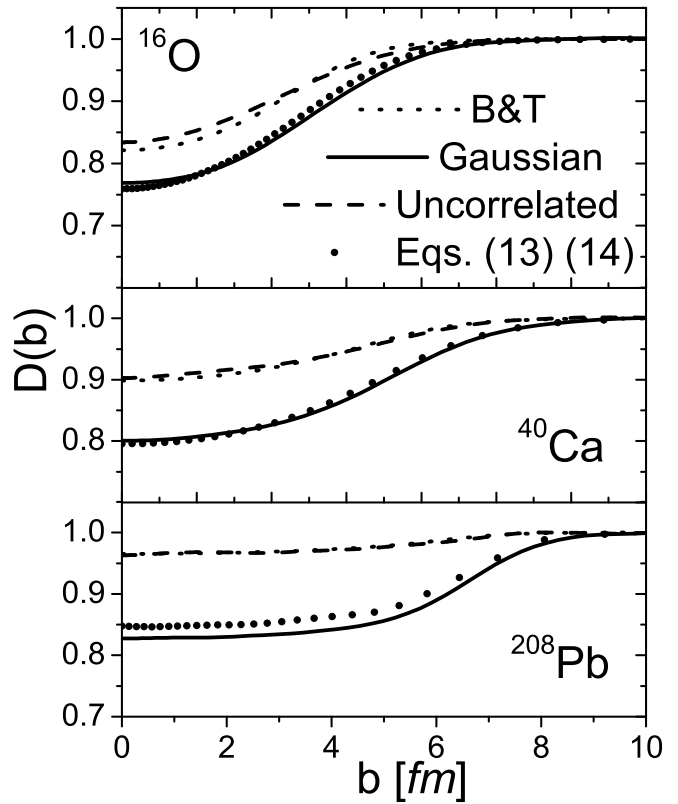}}
\vskip -0.3cm
\caption{The dispersion $D(b)$, as a function of the impact parameter $b$,
  defined in Eq.(\ref{aveDisp}) calculated for $^{16}O$, $^{40}Ca$ and
  $^{208}Pb$ nuclei, and for a $920$ $GeV$ incident nucleon.
  \textit{Dashed} curves correspond to uncorrelated configurations;
  \textit{solid lines} to the Gaussian correlations.
  The results obtained using eikonal model expression of B\&T (Ref.\cite{Bertocchi:1976bq})
  for the zero-range NN interactions are given by \textit{dotted} lines; for large
  A they coincide with the uncorrelated MC calculations.
  \textit{Symbols}: direct calculation using Eqs.(\ref{aveN}), (\ref{aveNNm1}),
  from Ref.\cite{Baym:1995cz}, with $C(r)$ extracted from the MC configurations and
  shown in Fig.\ref{Fig1}.}
\label{Fig4}
\end{figure}
%------------------------------------------------------------------------------%
%------------------------------------------------------------------------------%
the $^{16}O$ nucleus in  Fig.\ref{Fig2}, showing  a more  narrow distribution if
the approximation of Eq.(\ref{thickness}) is used; the same holds true for the other
considered nuclei.
The quantities $\langle N\rangle$ and $\langle N(N-1)\rangle$, can be written as
follows\cite{Baym:1995cz}:
\beqy
\label{aveN}
&&\hspace{-1.0cm}\langle N\rangle\,=\,\int d\Vec{r}_1\,\rho(\Vec{r}_1)
\,\frac{d\sigma^{in}}{d\Vec{b}}\,(\Vec{b}-\Vec{b}_1)\,,\\
&&\hspace{-1.0cm}\langle N(N-1)\rangle\,=\nonumber\\
\label{aveNNm1}&&\hspace{-0.5cm}
\int d\Vec{r}_1 d\Vec{r}_2
\,\rho^{(2)}(\Vec{r}_1,\Vec{r}_2)
\frac{d\sigma^{in}}{d\Vec{b}}(\Vec{b}-\Vec{b}_1)
\frac{d\sigma^{in}}{d\Vec{b}}(\Vec{b}-\Vec{b}_2)
\eeqy
which can be calculated in the present framework using
\beq
\label{dsigma}
\frac{d\sigma^{in}}{d\Vec{b}}\,(\Vec{b}-\Vec{b}_i)\,=\,1\,-
\,\left(1\,-\,\Gamma(\Vec{b}-\Vec{b}_i)\right)^2\,.
\eeq
Alternatively we can calculate $\langle N\rangle$, $\langle N(N-1)\rangle$
using Eqs.(\ref{aveN}), (\ref{aveNNm1}) as
\beq
\label{dispN1}
\langle N\rangle\,=\,\sum_N\,N\,P_N(b)\,,
\eeq
\beq
\label{dispN2}
\langle N(N-1)\rangle\,=\,\sum_N\,\left(N^2\,-\,N\right)\,P_N(b)\,.
\eeq
%------------------------------------------------------------------------------%
%------------------------------------------------------------------------------%
\begin{figure}[!h]
\vskip -0.0cm
\centerline{
    \includegraphics[width=9.5cm]{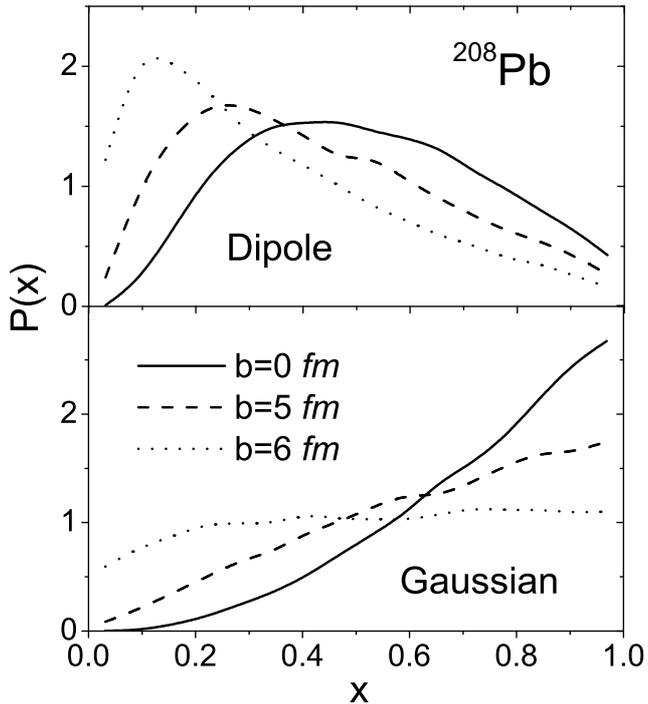}}
\vskip -0.3cm
\caption{The probability density $P(x)$ for the quantity $x$ defined in Eq.(\ref{piix})
    calculated as a function of the impact parameter $b$, for $m^2_g=0.6$ $GeV^2$ and
    for $^{208}Pb$. We have used the dipole (\textit{top}) and Gaussian (\textit{left})
    gluon density distributions of Ref.\cite{Frankfurt:2006jp} and correlated configurations.}
\label{Fig5}
\end{figure}
%------------------------------------------------------------------------------%
%------------------------------------------------------------------------------%
As a result we can write the variance of the mean
number of collisions as a function of the impact parameter $b$, as follows
\beqy
D(b)&=&\frac{\langle N^2\rangle\,-\,\left[\langle N\rangle\right]^2}{\langle N\rangle}
\nonumber\\
\label{aveDisp}&=&\frac{\sum_N\,N^2\,P_N(b)\,-\,\left[\sum_N\,N\,P_N(b)\right]^2}
{\sum_N\,N\,\,P_N(b)}\,.
\eeqy
In order to check the accuracy of our method, $D(b)$ was calculated both using Eqs.(\ref{aveN}),
(\ref{aveNNm1}) and Eq.(\ref{aveDisp}) using analytical one-body densities and $C(r)$ extracted
from the MC calculation; we then used Eqs.(\ref{dispN1}), (\ref{dispN2}), calculated with the $P_N(b)$
functions obtained with the MC. The results are compared in Fig.\ref{Fig4} for $^{16}O$, $^{40}Ca$
and $^{208}Pb$; a small discrepancy is exhibited, between the calculation with $C(r)$ (shown with
symbols) and the corresponding Gaussian correlated result, but the overall agreement is satisfactory.
The small difference between the results of the two methods, especially for lead, is to be ascribed
to the fact that the quantity $C(r)$ has been extracted from the MC configurations using Eq.(\ref{c12r}),
assuming the function $C(\Vec{r}_1,\Vec{r}_2)$ to be a function of the relative distance only, and
%------------------------------------------------------------------------------%
%------------------------------------------------------------------------------%
\begin{figure}[!ht]
\vskip -0.1cm
\centerline{
    \includegraphics[width=10cm]{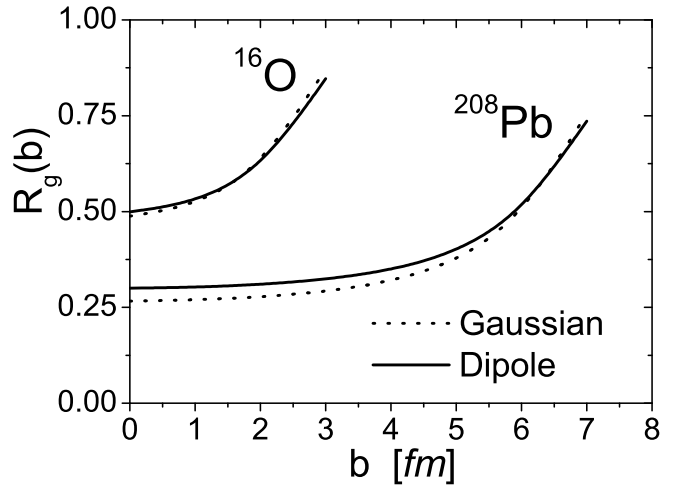}\hspace{0.4cm}}
\vskip -0.3cm
\caption{The ratio $R_g(b)$ defined in Eq.(\ref{ratb}) for $^{16}O$ and $^{208}Pb$,
    calculated using the gaussian (\textit{dots}) and dipole (\textit{solid}) gluon
    density distributions of Ref.\cite{Frankfurt:2006jp} and correlated configurations
    as a function of the impact parameter $b$, plotted
    for $m^2_g=0.6$ $GeV^2$ and for those values of $b$ for which there is enough
    statistics for the normalization $\int dx P(x)$ not to deviate from unity for
    more than $5\%$.}
\label{Fig6}
\end{figure}
%------------------------------------------------------------------------------%
%------------------------------------------------------------------------------%
not to depend on the particular region of the nucleus considered, while surface effects should be
present on the level of few \%.
We have compared the performance of the Glauber type interaction with the approach of
Ref.\cite{pie02}, where the probability of interaction in Eq.(\ref{probs}) was taken
as a step function, which vanishes for nucleons sitting outside the cylinder centered
at the given $|\Vec{b}|$ from the center of the nucleus. We took for the cylinder the
value which gives an area corresponding to the one given by
\beq
\hspace{-0.5cm}\sigma_{in}\,=\,\int d\Vec{b}\,\sigma_{in}(b)\,=
\,\int d\Vec{b}\,\left[1\,-\,
\left(1\,-\,\Gamma(\Vec{b}-\Vec{b}_i)\right)^2\right]\,
\eeq
where $\Gamma$ is calculated with the Glauber parameters we have used in the present work.
The results, as compared with our Glauber approach showed to differ substantially from both
the uncorrelated and correlated MC results.
\section{Lower limit on the parton  nuclear shadowing}
In this section  we present application of our MC code to the study of maximal possible nuclear
shadowing in nuclei. We use the well known observation that shadowing in the scattering off the
deuteron cannot reduce the cross section to the value smaller than cross section of scattering
off one nucleon. Basically it is due to one nucleon screening another one but not itself.

Similarly, it is natural to expect that in any  dynamics of parton interactions the gluon (quark)
density at a given impact parameter in a particular configuration cannot be less than the maximum
of individual transverse gluon densities, $g_N(x,\rho)$ in the nucleons at given b.
Here $g_N(x,\rho) = g_N(x) F_g(\rho)$ is the generalized diagonal gluon density in the nucleon.
%------------------------------------------------------------------------------%
%------------------------------------------------------------------------------%
\begin{figure}[!ht]
\vskip -0.0cm
\centerline{
    \includegraphics[width=10cm]{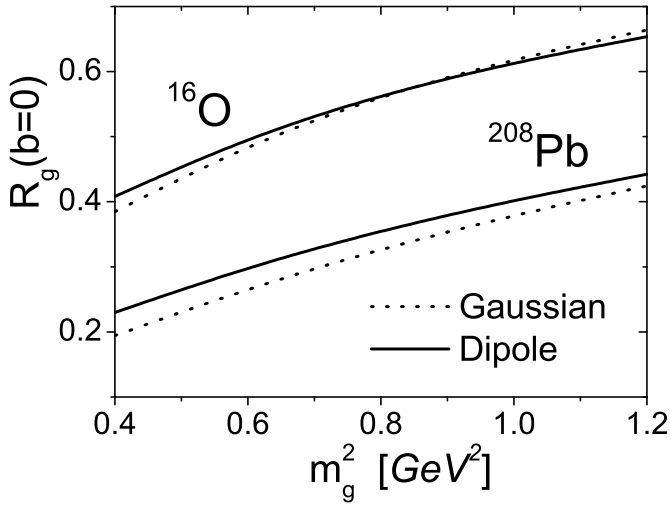}}
\vskip -0.3cm
\caption{The same quantity of Fig. \ref{Fig6} but plotted at $b=0$ and as
    a function of $m^2_g$, which is a parameter in Eq.(\ref{ratb}).}
\label{Fig7}
\end{figure}
%------------------------------------------------------------------------------%
%------------------------------------------------------------------------------%
\beq
g_A(x,b)_{conf} \ge max_{i=1, A} (g_N^{i}(x,r_t - b))\,,
\eeq
leading to
\beq
g_A(x,b)^{min} \ge \left <g_A(x,b)_{conf}\right>\,,
\eeq
In the limit of  $A\to \infty $ this leads to
 \begin{equation}
g_A(x,b) \ge g_N(\rho=0).
\end{equation}

The onset of the limiting behavior depends on the transverse shape of the gluon GPD. For the
same $g_N(x)$  a shape more peaked at small $\rho $ will asymptotically lead to larger value
of $g_A(x,b)^{min}$ though the approach to the  asymptotic value will require larger values
of $A$ as chances that there is a nucleon with $\rho$ small enough that $g_N(x,\rho)$ is close
to $g_N(x,0)$ is smaller in this case.
In Ref.\cite{Frankfurt:2006jp}, two parameterizations for the two-gluon form factor were
discussed, fitted to the $J/\psi$ photoproduction data (\cite{gluondata});
they were taken in the forms of an exponential and a dipole. The corresponding transverse
spatial distributions of gluon GPDs are thus
\beqy
F^1_g(\Vec{\rho})&=&\frac{1}{2\pi B_g}e^{-\rho^2/(2B_g)}\,,\nonumber\\
\label{gluondens}
F^2_g(\Vec{\rho})&=&\frac{m^2_g}{2\pi}\frac{m_g\rho}{2}K_1(m_g \rho)\,,
\eeqy
where $K_1$ is the modified Bessel function, $m^2_g=0.6$ $GeV^{2}$ for $x\sim 10^{-4}$
and $B_g=3.24/m^2_g$.
Using the configurations described in the previous sections, we have calculated for given
impact parameter $b$ the maximum transverse gluon density normalized to its peak value,
i.e. the maximum value of the probability density $P(x,b)$ of the quantity
\begin{equation}
\label{piix}
x=F_g(\Vec{\rho}_j)/F_g(0)\,,
\end{equation}
as a function of $\Vec{\rho}_j=\Vec{b}-\Vec{b}_j$, with $\Vec{b}_j$ the $j$-th nucleon
transverse coordinate; here $P(x,b)$ is normalized according to $\int dx P(b,x)=1$,
for given $b$.
%------------------------------------------------------------------------------%
%------------------------------------------------------------------------------%
\begin{figure}[!ht]
\vskip -0.0cm
\centerline{
    \includegraphics[width=10.2cm]{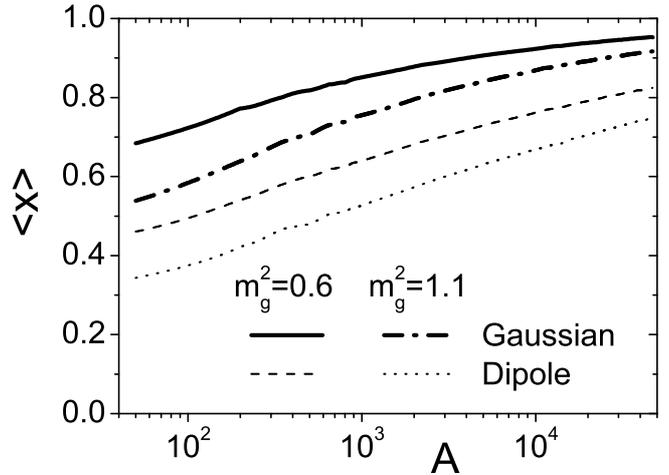}}
\vskip -0.3cm
\caption{The quantity $<x>$ as a function of $A$, calculated with the densities
    of Eqs.(\ref{gluondens}) and large values of $A$. The values of $m^2_g$
    are given in $GeV^2$.}
\label{Fig8}
\end{figure}
%------------------------------------------------------------------------------%
%------------------------------------------------------------------------------%
The results of the calculation for $^{208}Pb$ are shown in Fig.\ref{Fig5}. It can be
seen that the two densities of Eq.(\ref{gluondens}) produce very similar results.

We next apply these results to determine minimal value of the gluon shadowing at given
$b$ defined as the ratio of lower limit  on $g_A(x,b) $ and its value in the impulse
approximation, $g_A(x,b) =g_N(x)T_A(b)$; to this end, we take the ratio
\beqy
R_g(b)&=&\frac{g_N(x,\rho=0)\,\int dx\,x\,P(x,b)}{g_N(x)\,T_A(b)}\nonumber\\
\label{ratb}&=&\frac{F_g(0)\,\int dx\,x\,P(x,b)}{T_A(b)}\,.
\eeqy
Results are presented for the two used models of the gluon GPDs of Eq.(\ref{gluondens})
in Fig.\ref{Fig6} and Fig.\ref{Fig7}.
For $A\rightarrow\infty$ the limits using the two different fits differ by
$F^{(1)}_g(0)/F^{(2)}_g(0)=2/3.24=0.62$; however, this limit is reached at extremely
large A, as shown in Fig.\ref{Fig8}. This is due to a rather small radius of the
transverse gluon density $r^{tr}_g\le 0.5$ $fm$ and low nuclear density, leading
to a small probability for more than three nucleons to significantly screen each other
up to very large A. To investigate the onset of asymptotic, we evaluated the A dependence
of the quantity  $\langle x\rangle = \int dx P(x)$ at zero impact parameter for the two
considered gluon GPDs for two values of $m^2_g$ corresponding to $x\sim 0.01$ and $0.6$ $GeV^2$
corresponding to $x\sim 10^{-4}$. We modeled nuclei with very large A by
generating random nucleons with constant density $\rho_0=0.17$ $fm^{-3}$ in a cylinder
centered at $b=0$, with height $2 R_A$, where $R_A = 1.25 A^{1/3}$; the results, shown
in Fig.\ref{Fig8}, are independent from the radius of the cylinder, provided it is larger
than about $0.8$ $fm$. It can be seen that, the GPDs being very peaked at $\rho=0$,
the increase of $\langle x\rangle$ is very slow and the maximum value is not reached
even with A as large as $10^5$. Onset of asymptotic is somewhat faster for smaller $m_g$
due to a smoother behavior of $GPDs$ at $\rho\simeq 0$.
\section{Conclusions}
We developed a MC event generator for nucleon configurations in nuclei which
correctly reproduces single nucleon densities and central NN correlations in nuclei.
The generator can be used in modeling a wide range of processes. In particular,
it would be interesting to explore how it would modify effects of fluctuations
of the number of wounded nucleons in the heavy ion collisions.
\section{Acknowledgments}
We thank G. Baym, C. Ciofi degli Atti and L. Frankfurt for useful discussions.
\textit{This work is supported by DOE grant under contract DE-FG02-93ER40771}
%------------------------------------------------------------------------------%
%------------------------------------------------------------------------------%

%------------------------------------------------------------------------------%
%------------------------------------------------------------------------------%
%------------------------------------------------------------------------------%
%------------------------------------------------------------------------------%

\begin{thebibliography}{99}
  %
%\cite{Higinbotham:2009zz}
\bibitem{Higinbotham:2009zz}
  D.~Higinbotham, E.~Piasetzky and M.~Strikman,
  %``Protons and neutrons cosy up in nuclei and neutron stars,''
  CERN Cour.\  {\bf 49N1} (2009) 22
  %%CITATION = CECOA,49N1,22;%%
  %
%\cite{Subedi:2008zz}
\bibitem{Subedi:2008zz}
  R.~Subedi {\it et al.},
  %``Probing Cold Dense Nuclear Matter,''
  Science {\bf 320} (2008) 1476
  %%CITATION = SCIEA,320,1476;%%
  %
%\cite{Shneor:2007tu}
\bibitem{Shneor:2007tu}
  R.~Shneor {\it et al.}  [Jefferson Lab Hall A Collaboration],
  %``Investigation of Proton-Proton Short-Range Correlations via the 12C(e,e'pp)
  %Reaction,''
  Phys.\ Rev.\ Lett.\  {\bf 99} (2007) 072501;
  %%CITATION = PRLTA,99,072501;%%
  K. S. Egiyan at al. (CLAS), Phys. Rev. Lett. {\bf 96} (2006) 082501;
  A. Tang et al., Phys. Rev. Lett. {\bf 90} (2003) 042301
  %
%\cite{Frankfurt:2008zv}
\bibitem{Frankfurt:2008zv}
  L.~Frankfurt, M.~Sargsian and M.~Strikman,
  %``Recent observation of short range nucleon correlations in nuclei and their
  %implications for the structure of nuclei and neutron stars,''
  Int.\ J.\ Mod.\ Phys.\  A {\bf 23} (2008) 2991
  %%CITATION = IMPAE,A23,2991;%%
  %
  %\cite{Alvioli:2008rw}
\bibitem{Alvioli:2008rw}
  M.~Alvioli, C.~Ciofi degli Atti, I.~Marchino, V.~Palli and H.~Morita,
  %``Effects of Ground-State Correlations on High Energy Scattering off Nuclei:
  %the Case of the Total Neutron-Nucleus Cross Section,''
  Phys.\ Rev.\  C {\bf 78} (2008) 031601
  %%CITATION = PHRVA,C78,031601;%%
  %
  %\cite{Baym:1995cz}
\bibitem{Baym:1995cz}
  G.~Baym, B.~Blattel, L.~L.~Frankfurt, H.~Heiselberg and M.~Strikman,
  %``Correlations and fluctuations in high-energy nuclear collisions,''
  Phys.\ Rev.\  C {\bf 52} (1995) 1604
  %%CITATION = PHRVA,C52,1604;%%
  %
  %\cite{Alvioli:2005cz}
\bibitem{Alvioli:2005cz}
  M.~Alvioli, C.~Ciofi degli Atti and H.~Morita,
  %``Ground-state energies, densities and momentum distributions in closed-shell
  %nuclei calculated within a cluster expansion approach and realistic
  %interactions,''
  Phys.\ Rev.\  C {\bf 72} (2005) 054310
  %%CITATION = PHRVA,C72,054310;%%
  %
%\cite{Pandharipande:1979bv}
\bibitem{Pandharipande:1979bv}
  V.~R.~Pandharipande and R.~B.~Wiringa,
  Rev.\ Mod.\ Phys.\  {\bf 51} (1979) 821
  %\cite{Alvioli:2007zz}
\bibitem{Alvioli:2007zz}
  M.~Alvioli, C.~Ciofi degli Atti and H.~Morita,
  %``Proton-Neutron And Proton-Proton Correlations In Medium-Weight Nuclei And
  %The Role Of The Tensor Force,''
  Phys.\ Rev.\ Lett.\  {\bf 100} (2008) 162503
  %%CITATION = PRLTA,100,162503;%%
  %
%\cite{Schiavilla:2006xx}
\bibitem{Schiavilla:2006xx}
  R.~Schiavilla, R.~B.~Wiringa, S.~C.~Pieper and J.~Carlson,
  %``Tensor Forces and the Ground-State Structure of Nuclei,''
  Phys.\ Rev.\ Lett.\  {\bf 98} (2007) 132501
  %%CITATION = PRLTA,98,132501;%%
%  Phys. Rev. C {\bf 78} (2008) 021011
  %
  %\cite{Gyulassy:1994ew}
\bibitem{Gyulassy:1994ew}
  M.~Gyulassy and X.~N.~Wang,
  %``HIJING 1.0: A Monte Carlo program for parton and particle production in
  %high-energy hadronic and nuclear collisions,''
  Comput.\ Phys.\ Commun.\  {\bf 83} (1994) 307
  %%CITATION = CPHCB,83,307;%%
  %
  %\cite{Pieper:1992gr}
\bibitem{Pieper:1992gr}
  S.~C.~Pieper, R.~B.~Wiringa and V.~R.~Pandharipande,
  %``Variational calculation of the ground state of {$^{16}$}O,''
  Phys.\ Rev.\  C {\bf 46} (1992) 1741
  %%CITATION = PHRVA,C46,1741;%%
  %
\bibitem{download} \texttt{http://www.phys.psu.edu/\~{ }malvioli/eventgenerator}
  %
  \bibitem{alv04} M. Alvioli, C. Ciofi degli Atti, B. Z. Kopeliovich, I. K. Potashnikova,
  I. Schmidt
  {\it to be published}
  %
  %\cite{Bertocchi:1976bq}
\bibitem{Bertocchi:1976bq}
  L.~Bertocchi and D.~Treleani,
  %``Glauber Theory, Unitarity, And The Agk Cancellation,''
  J.\ Phys.\ G {\bf 3} (1977) 147
  %%CITATION = JPHGB,G3,147;%%
  %
  \bibitem{pie02} S. C. Pieper, in {\it Atomic and Nuclear Physics at One Gigaflop,
  Nuclear Science Research Conference Series} Volume 16, Oak Ridge, April 1988,
  edited by C. Bottcher, M. R. Stayer and J. B. McGrory.
  %
%\cite{Frankfurt:2006jp}
\bibitem{Frankfurt:2006jp}
  L.~Frankfurt, C.~E.~Hyde, M.~Strikman and C.~Weiss,
  %``Generalized parton distributions and rapidity gap survival in exclusive
  %diffractive $p p$ scattering,''
  Phys.\ Rev.\  D {\bf 75} (2007) 054009
  %%CITATION = PHRVA,D75,054009;%%
  %
\bibitem{gluondata} ZEUS Collaboration, S. Chekanov \textit{et al.},
  Eur. Phys. J. C {\bf 24} (2002) 345
  %
%  %\cite{Drescher:2006ca}
% \bibitem{Drescher:2006ca}
%   H.~J.~Drescher and Y.~Nara,
%   %``Effects of fluctuations on the initial eccentricity from the color  glass
%   %condensate in heavy ion collisions,''
%   Phys.\ Rev.\  C {\bf 75} (2007) 034905
%   %%CITATION = PHRVA,C75,034905;%%
  %
\end{thebibliography}
\end{document}